\documentclass{article}
\usepackage{amsfonts}

\usepackage{amsmath}
\usepackage{graphicx}

\begin{document}

\begin{center}

\bigskip\bigskip\bigskip

{\LARGE\textbf{MATTER SEEN AT MANY SCALES}}

\bigskip

{\LARGE\textbf{AND}}

\bigskip

{\LARGE\textbf{THE GEOMETRY OF AVERAGING}}

\bigskip

{\LARGE\textbf{IN RELATIVISTIC COSMOLOGY}}

\bigskip\bigskip\medskip

{\large Thomas Buchert
$\,^{a,}$\footnote{{\large {\small email: }  }
{\large {\small buchert@theorie.physik.uni-muenchen.de}}} 
$\;$and$\;$ Mauro Carfora $\,^{b,}$\footnote{{\large {\small email: }  }
{\large {\small mauro.carfora@pv.infn.it}}}} \vspace{24pt} \vspace{24pt}

$^{a}$~D\'epartement de Physique Th\'eorique, Universit\'e de Gen\`eve,\\[0pt%
]
24 quai E. Ansermet, CH--1211 Gen\`eve, Switzerland,\\
and\\[0pt]

Theoretische Physik, Ludwig--Maximilians--Universit\"at,\\
Theresienstr. 37, D--80333 M\"unchen, Germany

\medskip

$^{b}$~Dipartimento di Fisica Nucleare e Teorica,

Universit\`{a} degli Studi di Pavia, \\[0pt]

via A. Bassi 6, I--27100 Pavia, Italy, \\[0pt]

and\\[0pt]

Istituto Nazionale di Fisica Nucleare, Sezione di Pavia, \\[0pt]

via A. Bassi 6, I--27100 Pavia, Italy

\bigskip\medskip

\abstract{
We investigate the scale--dependence of Eulerian volume averages of scalar 
functions on Riemannian
three--manifolds. We propose a complementary view of a Lagrangian scaling of 
variables as opposed to their
Eulerian averaging on spatial domains. This program explains rigorously the 
origin of the
Ricci deformation flow for the metric, a flow which, on heuristic grounds, has 
been already suggested as a possible candidate for averaging the initial data 
set for cosmological spacetimes.}

\end{center}

\newpage

\rightline{\sl To Ruggiero, always open to learning new things}

\section{Introduction}

Averaged inhomogeneous cosmologies are in the forefront of interest, since
cosmological parameters like the rate of expansion or the mass density are
to be considered as volume--averaged quantities and only these can be
compared with observations 
\cite{buchert:grgdust}, \cite{buchert:onaverage}, \cite{buchert:average}, 
\cite{buchert:bks}, \cite{carfora:deformation1}, \cite{ellis:relativistic}, 
\cite{hosoya:RG}, \cite{sota:RG}. 
For this reason the relevant parameters are
intrinsically scale--dependent and one wishes to control this dependence
without restricting the cosmological model by unphysical assumptions. \
Consider a three--dimensional manifold equipped with a Riemannian metric $%
(\Sigma ,g)$ as a part of the initial data set for a cosmological
spacetime. On such a hypersurface we may select a simply--connected spatial
region and evaluate certain average properties of the physical
(scalar) variables on that domain such as, e.g., the volume--averaged
mass density field, or the volume--averaged scalar curvature. These (covariant)
average values are functionally dependent on position and shape of the
chosen domain of averaging. Let us now be more specific and relate the
averages to scaling properties of the physical variables. It is then natural
to identify the domain with a geodesic ball centred on a given position, and
introduce the spatial scale via the geodesic radius of the ball. We
may consider the variation of this radius and so explore the
scaling properties of average characteristics on the whole manifold. Doing
so for every position we arrive at smoothed fields that capture the
effective scale--dependent dynamics and so naturally provide the theoretical
values of observables that are drawn from a limited survey volume of a
cosmological spacetime.

\medskip

We may call this point of view Eulerian -- and everybody would first think
of this point of view -- in the sense that the spatial manifold is explored 
\emph{passively} by blowing up the geodesic balls covering larger and larger
volumes. On the contrary, the key idea of our approach consists in
demonstrating that a corresponding \emph{active} averaging procedure can be
devised which is Lagrangian: we hold the ball at a fixed (Lagrangian)
radius, and deform the dynamical variables inside the ball such that the deformation
corresponds to the smoothing of the fields. The variation with scale of, 
\emph{e.g.}, the density field or the metric is then mirrored by
one--parameter families of successively deformed density fields and metrics.
We shall show rigorously that this deformation corresponds to a first
variation of the metric along the Ricci tensor, known as the Ricci flow.
Since this flow has received great attention in the mathematical literature
-- major contributions are due to Hamilton \cite{hamilton:ricciflow1}, \cite
{hamilton:ricciflow2} -- we shall so translate the averaging procedure into
a well--studied deformation flow, much in the spirit of a renormalization
group flow \cite{carfora:deformation1}, \cite{carfora:deformation2}, \cite
{carfora:RG}, \cite{hosoya:RG}. This analysis, putting on rigorous grounds
the previous, largely heuristic use of the Ricci flow in smoothing out
relativistic spacetimes, is part of a larger program aimed to a full--fledged
analysis of averaging and scaling in relativistic cosmology. Such a program
features a mass--preserving Ricci flow and its variants in order to provide a
subtle technique for modelling an effective dynamics of inhomogeneities in
the Universe.

\section{Averaging and Scaling put into a Geometrical Perspective}

In order to characterize the correct conceptual framework for addressing
averaging and scaling properties in relativistic cosmology, let us remark
that in general relativity we basically have one scaling variable related to
the unit of distance: we can express the unit of time in terms of the unit
of distance using the speed of light. Similarly, we can express the unit of
mass through the unit of distance by using Newton's gravitational constant.
This remark implies that the scaling properties of Einstein's equations
typically generate a mapping between distinct initial data sets, (in this
sense we are basically dealing with a renormalization 
group transformation).
As we shall see below in detail, a rigorous characterization of an averaging
procedure in relativistic cosmology is indeed strictly connected to the
scaling geometry of the initial data set for Einstein's equations. 

\bigskip

In order to discuss the scaling properties and the averaging procedure
associated with an admissible set of initial data for a cosmological 
spacetime (we think of a setting within the
Arnowitt--Deser--Misner formulation of general relativity), 
a basic issue is to characterize
explicitly a scale--dependent averaging for the empirical mass distribution $%
\varrho $. This will be the main theme of this paper.

\subsection{Matter seen at many scales}

In order to characterize the scale over which we are smoothing the empirical
mass distribution $\varrho $, we need to study the distribution $\varrho $ by
looking at its average behavior on (geodesic) balls on $(\Sigma ,g)$ with
different centers and radii. The idea is to move from the function $\varrho
:\Sigma \rightarrow \mathbb{R}^{+}$ to an associated function, defined on $%
\Sigma \times \mathbb{R}^{+}$, and which captures some aspect of the
behavior of the given $\varrho $ on average, at different scales and locations.
The simplest function of this type is provided by the local volume average 
\begin{equation}
\left\langle \varrho \right\rangle _{B(p;r)}\doteq 
\frac{1}{Vol\left(
B(p;r)\right) }\int_{B(p;r)}\varrho d\mu _{g},  \label{locav}
\end{equation}
where $p\in \Sigma $ is a generic point, $d\mu _{g}$ is the Riemannian
volume element associated with $(\Sigma ,g)$,\ and $B(p;r)$ denotes the
geodesic ball at center $p$ and radius $r$ in $(\Sigma ,g)$, \emph{i.e.}, 
\begin{equation}
B(p;r)\doteq \left\{ q\in (\Sigma ,g)\,:d_{g}(p,q)\leq r\,\,\right\} ,
\end{equation}
where $d_{g}(p,q)$ denotes the distance, in $(\Sigma ,g)$, between the point 
$p$ and $q$. Note that if 
\begin{equation}
diam\doteq \sup \left\{ d_{g}(p,q)\,:p,q\in (\Sigma ,g)\right\}
\end{equation}
denotes the diameter of $(\Sigma ,g)$, then as $r\rightarrow diam$, we get 
\begin{equation}
\left\langle \varrho \right\rangle _{B(p;r)}\longrightarrow \left\langle \varrho
\right\rangle _{\Sigma }\doteq \frac{1}{Vol\left( (\Sigma ,g)\right) }%
\int_{\Sigma }\varrho d\mu _{g},
\end{equation}
at any point \bigskip\ $p\in \Sigma $. Conversely, if $\varrho :\Sigma
\rightarrow \mathbb{R}^{+}$ is locally summable then 
\begin{equation}
\lim_{r\rightarrow 0}\,\left\langle \varrho \right\rangle _{B(p;r)}=\varrho (p),
\end{equation}
for almost all points $p\in \Sigma $. The passage from $\varrho $ to $%
\left\langle \varrho \right\rangle _{B(p;r)}$ corresponds to replacing the
position--dependent empirical distribution of matter in $B(p;r)$ by a
locally uniform distribution $\left\langle \varrho \right\rangle _{B(p;r)}$
which is constant over the typical scale $r$. Note that for the total
(material) mass contained in $B(p;r)$, we get 
\begin{equation}
M(B(p;r))\doteq \int_{B(p;r)}\varrho d\mu _{g}=Vol\left( B(p;r)\right)
\left\langle \varrho \right\rangle _{B(p;r)},  \label{locmass}
\end{equation}
and 
\begin{equation}
\lim_{r\rightarrow diam}M(B(p;r))=M(\Sigma )\doteq \int_{\Sigma }\varrho d\mu _{g},
\end{equation}
where $M(\Sigma )$ is the total (material) mass contained in $(\Sigma ,g)$.
Our expectations in $\left\langle \varrho \right\rangle _{B(p;r)}$ are
motivated by the fact that on Euclidean $3$--space $(\mathbb{R}%
^{3},\delta _{ab})$, if $\varrho $ is a bounded function, then\ the local
average $\left\langle \varrho \right\rangle _{B(p;r)}$ is a Lipschitz function
on $\mathbb{R}^{3}\times \mathbb{R}^{+}$, \emph{i.e.}, \ 
\begin{equation}
\left| \left\langle \varrho \right\rangle _{B(p;r)}-\left\langle \varrho
\right\rangle _{B(q;r)}\right| \leq C_{0}d_{h}(p,q),
\end{equation}
where $C_{0}$ is a constant and $d_{h}(p,q)$ is the (hyperbolic!) distance
between $p$ and $q$. Thus, the local averages $\left\langle \varrho
\right\rangle _{B(p;r)}$ do not oscillate too wildly 
as $p$ and $r$ vary,
and the replacement of $\varrho $ by $\{\left\langle \varrho \right\rangle
_{B(p;r)}\}_{p\in \mathbb{R}^{3}}$ indeed provides an averaging of the
original matter distribution over the length scale $r$.\ It is not obvious
that such a nice behavior carries over to the Riemannian manifold $(\Sigma
,g)$. The point is that, even if the local averages $\{\left\langle \varrho
\right\rangle _{B(p;r)}\}_{p\in \Sigma }$ \ provide a controllable device of
smoothing the matter distribution at the given scale $r$, \ they still
depend in a sensible way on the geometry of the typical ball $B(p;r)$ as
we vary the averaging radius. In this connection we need to understand how,
as we rescale the domain $B(p;r)$, the local average $\left\langle \varrho
\right\rangle _{B(p;r)}$ depends on the underlying 
geometry of $(\Sigma ,g)$%
. The reasoning here is slightly delicate, so we go into a few details that
require some geometric preliminaries.

\bigskip

Let us denote by 
\begin{gather}
\exp _{p}:T_{p}\Sigma \rightarrow \Sigma  \label{expmap} \\
(\vec{v},r)\longmapsto \exp _{p}(r\vec{v})  \notag
\end{gather}
the exponential mapping at $p\in (\Sigma ,g)$, \emph{i.e.}, the map which
to the vector $r\vec{v}\in T_{p}\Sigma \simeq \mathbb{R}^{3}$
associates the point $\exp _{p}(r\vec{v})\in \Sigma $ reached \
at ``time'' $r\in \mathbb{R}_{+}$\ by the unique geodesic issued at $p\in
\Sigma $ with unit speed $\vec{v}\in \mathbb{S}^{2}(1)$ . Let \ $%
r>0$ be such that $\exp _{p}$ is defined on the Euclidean ball 
\begin{equation}
B_{E}(0,r)\doteq \{y\in T_{p}\Sigma \simeq \mathbb{R}^{3}:|y|\leq r\}
\end{equation}
and $\exp _{p}:B_{E}(0,r)\rightarrow B(p,r)\subset \Sigma $ is a
diffeomorphism onto its image. The largest radius $r$ for which this
happens, as $p$ varies in $\Sigma $, is the injectivity radius $inj_{M\text{ 
}}$\ of $(\Sigma ,g)$. Recall that a curve $c_{p,q}:\mathbb{R}\supset
I\rightarrow (\Sigma ,g)$ connecting the base point $p$ with the point $q$
is called a segment, if its length $L(c_{p,q})$ is such that $%
L(c_{p,q})=d_g (p,q)$, and if it is parametrized 
proportional to arclength. The
domain in $T_{p}\Sigma $ over which geodesics issued from $p$\ are segments (%
\emph{i.e.}, are distance--realizing) is called the segment domain and is
defined according to 
\begin{equation}
\overline{Seg(p)}\doteq \left\{ \vec{v}\in T_{p}\Sigma ;\exp
_{p}(r\vec{v}):[0,1]\rightarrow \Sigma \;is\;a\;segment\right\} .
\end{equation}
Note that $\overline{Seg(p)}$ is a closed star--shaped subset of $T_{p}\Sigma 
$, and $\Sigma =\exp _{p}\left( \overline{Seg(p)}\right) $. The exponential
map acts diffeomorphically over the interior $Seg(p)$ of $\overline{Seg(p)}$,  
and $\overline{Seg(p)}-Seg(p)$, the cut locus of $p$ in 
$T_{p}\Sigma $, is
a set whose image under $\exp _{p}$ has zero measure in $\Sigma $. In
particular, it follows that 
\begin{equation}
M(\Sigma )\doteq \int_{\Sigma }\varrho (y)d\mu _{g}=\int_{Seg(p)}\varrho (\exp
_{p})\exp _{p}^{\ast }(d\mu _{g}),
\end{equation}
where $\exp _{p}^{\ast }(d\mu _{g})$ denotes the pull--back of the Riemannian
measure $d\mu _{g}$ over $Seg(p)\subset T_{p}\Sigma $ under the exponential
mapping.

\bigskip

Let $B(p;r_{0})$ denote a given geodesic ball of radius $r_{0}<inj_{M}$, and for vector fields $X$, $Y$, and $Z$, in $(B(p;r),g)$ let 
$R(X,Y)Z=R^a_{\;\,bcd}X^cY^dZ^b$ be the corresponding curvature operator. Since $%
r:(B(p;r),g)\rightarrow \mathbb{R}$ is a distance function (\emph{i.e.}, $%
|\nabla r|\equiv 1$), the geometry of $B(p;r)$ can be described by the
following set of equations \cite{petersen} :
\begin{gather}
(\nabla _{\partial _{r}}S)(X)+S^{2}(X)=-R(X,\partial _{r})\partial _{r},
\label{shape} \\
(L_{\partial _{r}}g)(X,Y)=2g(S(X),Y),  \label{smetr} \\
\nabla _{\partial _{r}}S=L_{\partial _{r}}S,  \label{slie}
\end{gather}
where $\partial _{r}=\nabla r$ is
the gradient of $r$, $L_{\partial _{r}}g$ is the Lie derivative of the
metric in the radial direction $\partial _{r}$, and the shape tensor $%
S=\nabla ^{2}r$ is the Hessian of $r$. Such equations
follow from the Gauss--Weingarten relations applied to study the $r$%
--constant slices $U_{r}\doteq \{\exp _{p}(r\vec{v})\in \Sigma
:r=const.\}$ \ which are the images in $(B(p;r),g)$ of the standard
Euclidean $2$--spheres $\mathbb{S}^{2}(r)\subset T_{p}\Sigma \simeq \mathbb{R}%
^{3}$\footnote{
The Hessian $S=\nabla ^{2}r$\ is basically the second fundamental form of
the immersion $U_{r}\hookrightarrow (B(p;r),g)$. We use the equivalent
characterization of \emph{shape tensor }in order to avoid confusion with the
standard second fundamental form of use in relativity.}.
The shape tensor $S$ measures how the bidimensional metric $g^{(2)}$
induced on $U_{r}$ by the embedding in $(B(p;r),g)$ rescales as the radial
distance $r$ varies. Further properties of the equations (\ref{shape}) and (%
\ref{smetr}) that we need are best seen by using polar geodesic coordinates
and harmonic coordinates. Recall that normal exponential coordinates at $p$
are geometrically defined by 
\begin{gather}
\exp _{p}^{-1}:B(p;r)\rightarrow T_{p}\Sigma \simeq \mathbb{R}^{3} \\
q\longmapsto \exp _{p}^{-1}(q)=(y^{i}) , \notag
\end{gather}
where $(y^{i})$ are the Cartesian components of the velocity vector $\vec v 
\in $ $T_{p}\Sigma $ characterizing the geodesic segment from $p$ to $q$. 
Such coordinates are unique up to the chosen identification of $%
T_{p}\Sigma $ with $\mathbb{R}^{3}$. Since we are dealing with a radial
rescaling, for our purposes a suitable identification is the one associated
with the use of polar coordinates in $T_{p}\Sigma \simeq \mathbb{R}^{3}$. We
therefore introduce an orthonormal frame $\{e_{1},e_{2},e_{3}\}$ in $%
T_{p}\Sigma $ such that $e_{1}\doteq \partial _{r}$ and with $%
\{e_{2},e_{3}\} $ an orthonormal frame on the unit $2$--sphere $\mathbb{S}%
^{2}(1)\subset T_{p}\Sigma $. We can extend such vector fields radially to
the whole $T_{p}\Sigma $; we consider also the dual coframe $\{\theta
^{2},\theta ^{3}\}$ \ associated with $\{e_{2},e_{3}\}$. The introduction of
such a polar coordinate system in $T_{p}\Sigma $ is independent of the
metric $g$ and thus is ideally suited for discussing geometrically the $r$%
--scaling properties of $\left\langle \varrho \right\rangle _{B(p;r)}$. If we
pull--back to $T_{p}\Sigma $ the metric $g$ of $B(p;r)\subset \Sigma $, we
get 
\begin{equation}
g=dr^{2}+g(e_{A },e_{B })\theta ^{A }\theta ^{B },\;A
,B =1,2,  \label{polar}
\end{equation}
where the components $g_{A B}\doteq g(e_{A },e_{B })$ $%
=g^{(2)}(r,\vartheta ,\varphi )$ are functions of the polar coordinates $%
(\vartheta ,\varphi )$ in $T_{p}\Sigma \simeq \mathbb{R}^{3}$, associated
with the coframe $\{\theta ^{2},\theta ^{3}\}$. Note that such a local
representation of the metric holds throughout the local chart $(B(p;r),\exp
_{p}^{-1})$ and not just at $p$. \ In Cartesian coordinates in $T_{p}\Sigma
\simeq \mathbb{R}^{3}$ one recovers the familiar expression\footnote{Latin indices
run through $1,2,3$; we adopt the summation convention.}:
\begin{equation}
g_{ab}=\delta _{ab}-\frac{1}{3}R_{akbl}(p)y^{k}y^{l}+O\left(
|y|^{3}\right) .  \label{cartesian}
\end{equation}
In polar geodesic coordinates $(r,\vartheta ,\varphi )$ the equations (\ref
{shape}) and (\ref{smetr}) take the explicit form 
\begin{gather}
\partial _{r}S_{\;\,j}^{i}=-S_{\;\,k}^{i}S_{\;\,j}^{k}-
R_{\;\,rjr}^{i},  \label{peters} \\
\partial _{r}g_{ij}=2S_{\;\,i}^{k}g_{kj} ,  \notag
\end{gather}
where $R_{\;\,rjr}^{i}$ denotes the radial components of the curvature tensor.
By taking the trace of both equations we get 
\begin{gather}
\partial _{r}S=-S_{\;\,k}^{i}S_{\;\,i}^{k}-
Ric(\partial _{r},\partial _{r}),
\label{factor} \\
g^{ij}\partial _{r}g_{ij}=2S  , \notag
\end{gather}
where $Ric(\partial _{r},\partial _{r})\doteq $ 
$R_{\;\,rir}^{i}$\ denotes the $%
\partial _{r}$ component of the Ricci tensor, and $S\doteq S_{\;\,k}^{i}\delta
_{\;\,i}^{k}$ is the rate of volume expansion of 
\ $g^{(2)}(r,\vartheta ,\varphi
)$, (the first equation in (\ref{factor}) is nothing but the Jacobi operator
coming from the second variation of the area for $g^{(2)}(r,\vartheta
,\varphi )$). If we assume that the curvature $R(X,Y)Z$ 
shall then be given by fixing
the $(\vartheta ,\varphi )$--dependence in the factorized 
metric (\ref{polar}%
), we can consider (\ref{peters}) as a system of decoupled ordinary
differential equations\ describing \ the rescaling of the geometry of \ $%
B(p;r)$ in terms of the one--parameter flow of \ immersions $\mathbb{S}%
^{2}(r)\mapsto $ $(U_{r},g^{(2)}(r,\vartheta ,\varphi ))$. Note in
particular that the shape tensor matrix $(S_{\;\,j}^{i})$ is characterized, 
from
the first set of equations (\ref{peters}), as a functional of the given
curvature tensor $(R_{\;\,rjr}^{i})$. In such a setting the equation $\partial
_{r}g_{ij}=2S_{\;\,i}^{k}g_{kj}$ can be interpreted by saying that the metric
rescales radially along the (curvature--dependent) shape tensor 
$S_{\;\,i}^{k}$.
In a sufficiently small neighborhood of the pole $p$, from the expression (%
\ref{cartesian}) and from $\partial _{r}=\frac{1}{r}y^{i}\partial _{i}$, we
get that in Cartesian coordinates we can write 
\begin{equation}
\partial _{r}g_{ab}=-\frac{2}{9}rR_{ab}(p)-\frac{2}{3}\left[ \frac{y^{i}y^{k}%
}{r}-\frac{1}{3}\delta ^{ik}r\right] R_{aibk}(p)+O(r^{2}),  \label{radscale}
\end{equation}
where we have evidentiated the trace--free part 
$[y^{i}y^{k}-\frac{1}{3}%
\delta ^{ik}r^{2}]$ of the Cartesian tensor $y^{i}y^{k}$. \ Roughly
speaking, the term $\frac{2}{3}\left[ \frac{x^{i}x^{k}}{r}-\frac{1}{3}\delta
^{ik}r\right] R_{aibk}(p)$ represents the shear part of the rescaling of the
geometry of the spherical surfaces $U_{r}\doteq \{\exp _{p}(r\vec{%
v})\in \Sigma :r=const.\}$ as $r$ varies in the neighborhood of the pole $p$%
. The term $-\frac{2}{9}rR_{ab}(p)$ is instead responsible
for the radial
rescaling in the geometry of $U_{r}$. Thus, one may say that in a
sufficiently small neighborhood of the point $p$ the metric rescales
radially in the \emph{direction} of its Ricci tensor. 
This latter remark
will turn out quite useful in understanding the geometric rationale
behind the choice of a proper averaging procedure for the geometry of \ $%
(\Sigma ,g)$.

\bigskip

Even if polar geodesic coordinates suggest themselves as the most natural
labels for the points of \ $B(p;r)$, they suffer from the basic drawback
that their domain of definition cannot be a priori estimated and strongly
depends on the local geometry of $(\Sigma ,g)$. In this connection, a much
better control on the geometry of the balls $B(p;r)$, and hence on $%
\left\langle \varrho \right\rangle _{B(p;r)}$, can be achieved by labelling the
points $\exp _{p}^{-1}(q)\in $\ $B_{E}(0,r)$ with harmonic coordinates,
i.e., a coordinate system $\{X^{i}\}$ such that the coordinate functions $X^{i}$ are
harmonic functions with respect to the Laplacian on $(\Sigma ,g)$. We can do
this by starting from the given (Cartesian) normal coordinates $\{y^{k}\}$, and look for
a diffeomorphism on a sufficiently small Euclidean ball $B_{E}(0,r)\subset 
\mathbb{R}^{3}$, 
\begin{gather}
\Phi:B_{E}(0,r)\rightarrow B_{E}(0,r) \\
y^{k}=\exp _{p}^{-1}(q)\longmapsto \Phi (y^{k})\doteq X^{i}  \notag
\end{gather}
such that 
\begin{equation}
\left\{ 
\begin{tabular}{l}
$\Delta \Phi ^{k}=\frac{1}{\sqrt{\det (g_{ab})}}\partial _{i}\left( \sqrt{%
\det (g_{ab})}g^{ij}\partial _{j}\Phi ^{k}\right) =0$ \\ 
$\Phi ^{k}|_{\partial B_{E}(0,r)}=Id\;.$%
\end{tabular}
\right.
\end{equation}
The standard theory of elliptic partial differential equations\ implies that
the harmonic functions so characterized do form a coordinate system in $%
B_{E}(0,r)$. The important observation is that such harmonic coordinates can
be introduced on balls of an a priori size as soon as the manifold $(\Sigma
,g)$ has bounded sectional curvature and its injectivity radius is bounded
below. Note that from $\exp _{p}^{-1}:B(p;r)\rightarrow T_{p}\Sigma $ and a
generic (\emph{i.e.}, not necessarily harmonic) diffeomorphism $\Phi
:B_{E}(0,r)\rightarrow B_{E}(0,r)$ we can define the map 
\begin{equation}
F\doteq \Phi \circ \exp _{p}^{-1}:(B(p;r),g)\overset{\exp _{p}^{-1}}{%
\rightarrow }(B_{E}(0,r),\delta )\overset{\Phi }{\rightarrow }%
(B_{E}(0,r),\delta ),
\end{equation}
where $\delta $ denotes the Euclidean metric on 
$B_{E}(0,r)$. To any such
map we can associate the scalar ``energy density'' \cite{hamilton:ricciflow2}
\begin{equation}
e(x)\doteq \left\| dF\right\| ^{2}=g^{ij}\delta _{lm}\frac{\partial F^{l}}{%
\partial x^{i}}\frac{\partial F^{m}}{\partial x^{j}},  \label{endens}
\end{equation}
where the $\{x^{i}\}$ denote given generic coordinates of the points $q$\
of $(B(p;r),g)$ \ and where 
\begin{equation}
dF=\frac{\partial F^{l}}{\partial x^{i}}dx^{i}\otimes \frac{\partial }{%
\partial F^{l}}
\end{equation}
is the differential of the map $F$, \ thought of as a section of the bundle 
\begin{equation}
\left. T^{\ast }\Sigma \otimes F^{-1}T\mathbb{R}^{3}\right| _{B(p;r)} .
\end{equation}
Note that the pull--back bundle $F^{-1}T\mathbb{R}^{3}$ 
over $B(p;r)$
inherits the flat metric $\delta _{lm}(F(x))$. 
Thus $\left. T^{\ast }\Sigma
\otimes F^{-1}T\mathbb{R}^{3}\right| _{B(p;r)}$ is naturally endowed with
the metric $g^{ij}$ on $T^{\ast }\Sigma $ and $\delta _{lm}$ on $F^{-1}T%
\mathbb{R}^{3}$, a metric which characterizes the norm $\left\| dF\right\|
^{2}$. Alternatively, $e(x)$ is the trace with respect to the metric of $%
T^{\ast }\Sigma $ \ of the pull--back by $F$ of the flat metric tensor $%
(B_{E}(0,r),\delta )$. Such remarks emphasize the fact 
that, even if the
value of $e(x)$ seems to depend on the choice of local coordinates, \ it
is indeed a well--known geometrical quantity: the harmonic map energy density
associated with the map $F:(B(p;r),g)\rightarrow (B_{E}(0,r),\delta )$. As
to its meaning in our setting, note that we can always choose the given
generic coordinates $\{x^{i}\}$ to be the original geodesic coordinates $\{y^{i}\}$. By
exploiting such a remark, from the asymptotic expansion (\ref{cartesian}),
we easily get that in a sufficiently small neighborhood of the point $p$,
the scalar $e(x)$ can be written as 
\begin{equation}
e(y)=\left( \delta ^{ab}+\frac{1}{3}%
R^{a\;\,b}_{\;\,k\;\,l}(p)y^{k}y^{l}+O(|y|^{2})\right) \delta _{rs}\frac{\partial
F^{r}}{\partial y^{a}}\frac{\partial F^{s}}{\partial y^{b}}.
\end{equation}
If we further assume that the diffeomorphism $\Phi :B_{E}(0,r)\rightarrow
B_{E}(0,r)$ is the identity, (\emph{i.e.}, $F$ reduces to $\exp _{p}^{-1}$),
then $\frac{\partial F^{r}}{\partial y^{a}}=\delta _{\;\,a}^{r}$, and we get 
\begin{equation}
\left. e(y)\right| _{F=\exp _{p}^{-1}}=3+\frac{1}{3}%
R_{kl}(p)y^{k}y^{l}+O(|y|^{2})  \label{emeaning}
\end{equation}
showing that $\left. e(x)\right| _{F=\exp _{p}^{-1}}$ is, roughly speaking,
a measure of the curvature\emph{\ }associated with $(B(p;r),g;\exp
_{p}^{-1}) $. As we shall see, the quantity $e(x)$ will play a basic role in
understanding the geometric rationale for a proper averaging strategy of the
geometry of $(\Sigma ,g)$.

\bigskip

\bigskip

\ Guided by such geometrical features of geodesic balls, let us go back to
the study of the scaling properties of \ \ $\left\langle \varrho \right\rangle
_{B(p;r)}$. To this end, for any $r$ and $s$ such that $r+s<inj_{\Sigma }$,
let us consider the one--parameter family of diffeomorphisms (geodesic ball
dilatations) 
\begin{gather}
H_{s}:(\Sigma ,p)\rightarrow (\Sigma ,p) \\
q=\exp _{p}[r_{q}(\partial _{r},e_{2},e_{3})]\longmapsto H_{q}(q)\doteq \exp
_{p}[(r_{q}+s)(\partial _{r},e_{2},e_{3})]  \notag
\end{gather}
defined by flowing each point $q\in B(p;r)$ a distance $s$ along the
unique radial geodesic segment issued at $p\in \Sigma $ and passing \
through $q$. Let us remark that, for any $r$ such that $r_{0}\leq r<inj_{M}$%
, we can formally write \ 
\begin{equation}
B(p;r)=H_{(r-r_{0})}B(p;r_{0}).
\end{equation}
Thus, in a neighborhood of $B(p;r_{0})$ and for sufficiently small $r$, we
get 
\begin{equation}
M(B(p;r))\doteq \int_{B(p;r)}\varrho d\mu
_{g}=\int_{B(p;r_{0})}H_{(r-r_{0})}^{\ast }(\varrho d\mu _{g}) , \label{pullM}
\end{equation}
where $H_{(r-r_{0})}^{\ast }d(\varrho \mu _{g})$ is the Riemannian measure
obtained by pulling back $\varrho d\mu _{g}$ under the action of $H_{(r-r_{0})}$%
. By differentiating (\ref{pullM}) with respect to $r$, we have 
\begin{gather}
\frac{d}{dr}M(B(p;r))=\lim_{h\rightarrow 0}\frac{\left[ M(B(p;r+h))-M(B(p;r))%
\right] }{h} \\
=\lim_{h\rightarrow 0}\frac{\left[ \int_{B(p;r_{0})}H_{(r-r_{0})+h}^{\ast
}(\varrho d\mu _{g})-\int_{B(p;r_{0})}H_{(r-r_{0})}^{\ast }(\varrho d\mu _{g})%
\right] }{h}.  \notag
\end{gather}
Since $H_{(r-r_{0})+h}=H_{(r-r_{0})}\circ H_{h}$, we can write the above
expression as 
\begin{gather}
\lim_{h\rightarrow 0}\left[ \int_{B(p;r_{0})}\frac{H_{(r-r_{0})}^{\ast }%
\left[ H_{h}^{\ast }(\varrho d\mu _{g})-(\varrho d\mu _{g})\right] }{h}\right] \\
=\lim_{h\rightarrow 0}\left[ \int_{B(p;r)}\frac{\left[ H_{h}^{\ast }(\varrho
d\mu _{g})-(\varrho d\mu _{g})\right] }{h}\right]  \notag \\
=\int_{B(p;r)}\lim_{h\rightarrow 0}\frac{\left[ H_{h}^{\ast }(\varrho d\mu
_{g})-(\varrho d\mu _{g})\right] }{h}\;,  \notag
\end{gather}
from which it follows that 
\begin{gather}
\frac{d}{dr}M(B(p;r))=\int_{B(p;r)}L_{\partial _{r}}(\varrho d\mu _{g})
\label{demme} \\
=\int_{B(p;r)}\left( \frac{\partial }{\partial r}\varrho +\varrho div(\partial
_{r})\right) d\mu _{g}  \notag \\
=\int_{B(p;r)}\left( \frac{\partial }{\partial r}\varrho +\frac{1}{2}\varrho g^{ab}%
\frac{\partial }{\partial r}g_{ab}\right) d\mu _{g},  \notag
\end{gather}
where $L_{\partial _{r}}$ and $div(\partial _{r})$ denote the Lie derivative
along the vector field $\partial _{r\text{ \ }}$and its divergence,
respectively, and where we have exploited the well--known expression for the
derivative of a volume density, 
\begin{equation}
\frac{\partial }{\partial r}\sqrt{g}=\frac{1}{2}\sqrt{g}g^{ab}\frac{\partial 
}{\partial r}g_{ab}.
\end{equation}

With these preliminary remarks along the way, it is straightforward to
compute the rate of variation with $r$ of the\ local average $\left\langle
\varrho \right\rangle _{B(p;r)}$, since (\ref{demme}) implies 
\begin{gather}
\frac{d}{dr}\left\langle \varrho \right\rangle _{B(p;r)}=\left\langle \frac{%
\partial }{\partial r}\varrho \right\rangle _{B(p;r)}  \label{der} \\
+\frac{1}{2}\left\langle \varrho g^{ab}\frac{\partial }{\partial r}%
g_{ab}\right\rangle _{B(p;r)}-\frac{1}{2}\left\langle \varrho \right\rangle
_{B(p;r)}\left\langle g^{ab}\frac{\partial }{\partial r}g_{ab}\right\rangle
_{B(p;r)},  \notag
\end{gather}
where $\left\langle f\right\rangle _{B(p;r)}$ denotes the volume average of $%
f$ over the ball $B(p;r)$. Explicitly, by exploiting (\ref{peters}), we get 
\begin{equation}
\frac{d}{dr}\left\langle \varrho \right\rangle _{B(p;r)}=\left\langle \frac{%
\partial }{\partial r}\varrho \right\rangle _{B(p;r)}+\left\langle \varrho
S\right\rangle _{B(p;r)}-\left\langle \varrho \right\rangle
_{B(p;r)}\left\langle S\right\rangle _{B(p;r)}.
\end{equation}
Thus, the local average $\left\langle \varrho \right\rangle _{B(p;r)}$ feels
the fluctuations in the geometry as we vary the scale, fluctuations
represented by the shape tensor terms 
\begin{equation}
\left\langle \varrho S\right\rangle _{B(p;r)}-\left\langle \varrho \right\rangle
_{B_{E}(0;r)}\left\langle S\right\rangle _{B(p;r)}
\end{equation}
governed by the curvature in $B(p;r)$ according to (\ref{factor}), and\
expressing a geometric non--commutativity between averaging over the ball $%
B(p;r)$ and rescaling its size. Since the curvature varies both in the given 
$B(p;r)$ and when we consider distinct base points $p$, \ the above remarks
indicate that the local averages $\left\langle \varrho \right\rangle _{B(p;r)}$
\ are subjected to the accidents of the fluctuating geometry of $(\Sigma ,g)$%
. In other words,\ there is no way of obtaining a proper smoothing of $\varrho $
without smoothing out at the same time the geometry of $(\Sigma ,g)$.

\subsection{Eulerian averaging and Lagrangian scaling}

The use of the exponential mapping in discussing the geometry behind the
local averages $\left\langle \varrho \right\rangle _{B(p;r)}$ makes it clear
that we are trying to measure how different the \ $\left\langle \varrho
\right\rangle _{B(p;r)}$ are from the standard average over Euclidean balls.
In so doing we think of \ $\exp _{p}:T_{p}\Sigma \rightarrow \Sigma $ as
maps from the fixed space $B_{E}(0,r)$ into the manifold $(\Sigma ,g)$. In
this way we are implicitly trying to transfer information from the manifold $%
(\Sigma ,g)$ into domains of $\mathbb{R}^{3}$ which we would like to be, as
far as possible, to be independent of the accidental geometry of $(\Sigma
,g)$ itself. Indeed, any averaging would be quite difficult to implement if
the reference model varies with the geometry to be averaged. As already
emphasized, this latter task is only partially accomplished by the
exponential mapping, since the domain over which $\exp _{p}:T_{p}\Sigma
\rightarrow \Sigma $ is a diffeomorphism depends on $p$ and on the actual
geometry of $(\Sigma ,g)$. We can go a step further in this direction by
considering not just the given $(\Sigma ,g)$ but rather a whole family of
Riemannian manifolds.\ To start with let us remark that, if we fix the
radius $r_{0}$ of the Euclidean ball $B_{E}(0;r_{0})\subset T_{p}\Sigma $
and consider the family of exponential mappings, \ $\exp _{(p,\beta
)}:T_{p}\Sigma \rightarrow (\Sigma ,g(\beta ))$, associated with a
corresponding one--parameter family of \ Riemannian metrics $g_{ab}(\beta)$, 
$0\leq \beta <+\infty $, with $g_{ab}(\beta =0)=g_{ab}$, then $B(p;r_{0})$
becomes a functional of the set of \ Riemannian structures associated with $%
g(\beta )_{ab}$, $0\leq \beta <+\infty $,\emph{\ i.e.}, 
\begin{equation}
B(p;r_{0})\longmapsto B_{\beta }(p;r_{0})\doteq \exp _{(p,\beta )}\left[
B_{E}(0;r_{0})\right] .
\end{equation}

\ In this way, instead of considering just a given geodesic ball $B(p;r_{0})$%
, we can consider, as $\beta $ varies, a family of geodesic balls \ $%
B_{\beta }(p;r_{0})$, all with the same radius $r_{0}$ but with distinct
inner geometries $g_{ab}(\beta)$. Since $B_{\beta =0}(p;r_{0})=B(p;r_{0})$,
such balls can be thought of as obtained from the given one $B(p;r_{0})$ by a
smooth continuous deformation of its original geometry.\ Under such
deformation also $\ \left\langle \varrho \right\rangle _{B(p;r_{0})}$ becomes a
functional, $\left\langle \varrho \right\rangle _{B_{\beta }(p;r_{0})}$, of the
one--parameter family of \ Riemannian structures associated with 
$g_{ab}(\beta)$, $0\leq \beta <+\infty $. The elementary but basic observation in
order to take properly care of the geometrical fluctuations in $\
\left\langle \varrho \right\rangle _{B(p;r_{0})}$ is that the right member of (%
\ref{der}) has precisely the formal structure of the linearization (\emph{%
i.e.}, of the variation) of the functional $\left\langle \varrho \right\rangle
_{B(p;r_{0})}$ in the direction of \ the (infinitesimally) deformed
Riemannian metric \ $\partial / \partial\beta  \lbrack g_{ab}(\beta)\rbrack$, 
\emph{viz.}, 
\begin{gather}
\frac{d}{d\beta }\left\langle \varrho \right\rangle _{B_{\beta
}(p;r_{0})}=\left\langle \frac{\partial }{\partial \beta }\varrho \right\rangle
_{B_{\beta }(p;r_{0})}  \notag \\
+\left\langle \varrho \,g^{ab}(\beta )\frac{\partial }{\partial \beta }%
g_{ab}(\beta )\right\rangle _{B_{\beta }(p;r_{0})}-\left\langle \varrho
\right\rangle _{B_{\beta }(p;r_{0})}\left\langle g^{ab}(\beta )\frac{%
\partial }{\partial \beta }g_{ab}(\beta )\right\rangle _{B_{\beta
}(p;r_{0})},  \label{geodef}
\end{gather}
where the ball $B_{E}(0;r_{0})$ is \emph{kept fixed} while its image $%
B(p;r_{0})$ is deformed according to the flow of metrics $g_{ab}(\beta)$, $%
0\leq \beta \leq \infty $.

In a rather obvious sense, (\ref{geodef})\ represents the \emph{active}
interpretation corresponding to the Eulerian \emph{passive} view associated
with the ball variation $B(p;r_{0})\rightarrow B(p;r)$. In other words, we
are here dealing with the Lagrangian point of view of following a fluid
domain in its deformation, where the \emph{fluid particles} here are the
points of $B(p;r_{0})$ suitably labelled. This latter remark suggests that
in order to optimize the local averaging procedure associated with the local
average $\left\langle \varrho \right\rangle _{B(p;r)}$, instead of studying its
scaling behavior as $r$ increases, and consequently be subjected to the
accidents of the fluctuating geometry of $(B(p;r),g_{ab})$, we may keep
fixed the domain $B(p;r_{0})$ and rescale the geometry inside $B(p;r_{0})$
according to a suitable flow of metrics $g_{ab}\rightarrow g_{ab}(\beta)$, $%
0\leq \beta \leq \infty $. Correspondingly, also the matter density $\varrho $
will be forced to rescale $\varrho \rightarrow \varrho (\beta )$, and if we are
able to choose the flow $g_{ab}\rightarrow g_{ab}(\beta )$ in such a way
that the local inhomogeneities of the original geometry of $(\Sigma ,g)$ are
smoothly eliminated, then the local average $\left\langle \varrho \right\rangle
_{B(p;r)}$ comes closer and closer to represent a matter averaging over a
homogeneous geometry.

\bigskip

In order to put such Lagrangian picture on a firmer ground we need to
consider the important issue of how to label the points of $%
(B(p;r_{0}),g(\beta ))$ during the deformation process. For $\beta =0$ the
natural choice are (polar) geodesic coordinates associated
with $\exp _{p}$. However, as the metric changes along the flow $%
(B(p;r_{0}),g(\beta ))$, the remarks of the previous section suggest that
polar geodesic coordinates may not be the optimal choice as $\beta $ varies,
since we do not have an a priori control on the domain of validity of such a
coordinate system. However, if we select the flow $g_{ab}\rightarrow 
g_{ab}(\beta)$, $0\leq \beta \leq \infty $, in such a way that the sectional
curvatures of $g(\beta )$ stay bounded and the injectivity radius is bounded
below as $\beta $ varies, then such an a priori control can be achieved by
labelling the points of $\exp _{(p,\beta )}^{-1}$ by harmonic coordinates.
Thus, in order to properly label the points of $(B(p;r_{0}),g(\beta ))$,
instead of considering the one--parameter family of maps $\exp _{(p,\beta
)}^{-1}:(B(p;r_{0}),g(\beta ))\rightarrow (B_{E}(0;r_{0}),\delta )$, we
shall consider the family of maps 
\begin{equation}
F(\beta )\doteq \Phi (\beta )\circ \exp _{(p,\beta
)}^{-1}:(B(p;r_{0}),g(\beta ))\overset{\exp _{(p,\beta )}^{-1}}{\rightarrow }%
(B_{E}(0;r_{0}),\delta )\overset{\Phi (\beta )}{\rightarrow }%
(B_{E}(0;r_{0}),\delta ),
\end{equation}
where $\Phi (\beta ):(B_{E}(0;r_{0}),\delta )\rightarrow
(B_{E}(0;r_{0}),\delta )$ is a $\beta $--dependent diffeomorphism on $%
(B_{E}(0;r_{0}),\delta )$, (inducing a coordinate change in $%
(B_{E}(0;r_{0}),\delta )$). By mimicking (\ref{endens}), we associate the
harmonic map energy density with the family of maps $F(\beta )$ providing
the Lagrangian coordinates that allow to follow the points of $%
(B(p;r_{0}),g(\beta ))$ during the deformation,\emph{\ i.e.}, 
\begin{equation}
e(\beta ;x)\doteq \left\| dF(\beta )\right\| ^{2}=g^{ij}(\beta )\delta _{lm}%
\frac{\partial F^{l}(\beta )}{\partial x^{i}}\frac{\partial F^{m}(\beta )}{%
\partial x^{j}},
\end{equation}
where the $\{x^{i}\}$ denote the fixed generic (Eulerian) coordinates of
the points of $(\Sigma _{p},g(\beta ))$. If we compute the rate of variation
with $\beta $ of $e(\beta ;x)$ we get 
\begin{gather}
\frac{\partial }{\partial \beta }e(\beta ;x)=\delta _{lm}\frac{\partial
F^{l}(\beta )}{\partial x^{i}}\frac{\partial F^{m}(\beta )}{\partial x^{j}}%
\frac{\partial }{\partial \beta }g^{ij}(\beta )  \label{Eequat} \\
+2g^{ij}(\beta )\delta _{lm}\frac{\partial F^{l}(\beta )}{\partial x^{i}}%
\frac{\partial }{\partial x^{j}}\left( \frac{\partial }{\partial \beta }%
F^{m}(\beta )\right) .  \notag
\end{gather}
Since we eventually want to label the points of \ $(B(p;r_{0}),g(\beta ))$
with harmonic coordinates, the most natural thing to do is to \ deform $%
F(\beta )$ according to the harmonic map flow, \emph{i.e.}, 
\begin{equation}
\left\{ 
\begin{tabular}{l}
$\frac{\partial }{\partial \beta }F^{m}(\beta )=\Delta _{\beta }F^{m}(\beta
) $ \\ 
$F^{m}(\beta =0)=\exp _{p}^{-1}$ ,
\end{tabular}
\right.  \label{Eells}
\end{equation}
where 
\begin{equation}
\Delta _{\beta }F^{m}(\beta )\doteq g^{ij}(\beta )\left[ \frac{\partial ^{2}%
}{\partial x^{i}\partial x^{j}}F^{m}(\beta )-\Gamma _{\;\,ij}^{k}(g(\beta ))%
\frac{\partial }{\partial x^{k}}F^{m}(\beta )\right]
\end{equation}
is the harmonic map Laplacian associated with the Riemannian connection $%
\Gamma _{\;\,ij}^{k}(g(\beta ))$ of $(B(p;r_{0}),g(\beta ))$. In this way $%
F^{m}(\beta )$ evolves towards harmonic maps (i.e., the Lagrangian
coordinates labelling the points \ of $(B(p;r_{0}),g(\beta ))$ will be
harmonic coordinates as $\beta \rightarrow \infty $). Note that (\ref{Eells}%
) is a strictly parabolic initial value problem admitting a solution for
sufficiently small $\beta >0$, (in our case, standard theorems in harmonic
map theory imply that the solution actually exists for all $0\leq \beta
<\infty $). If we insert (\ref{Eells}) into (\ref{Eequat}) we get 
\begin{gather}
\frac{\partial }{\partial \beta }e(\beta ;x)=\delta _{lm}\frac{\partial
F^{l}(\beta )}{\partial x^{i}}\frac{\partial F^{m}(\beta )}{\partial x^{j}}%
\frac{\partial }{\partial \beta }g^{ij}(\beta )  \label{ener2} \\
+2g^{ij}(\beta )\delta _{lm}\frac{\partial F^{l}(\beta )}{\partial x^{i}}%
\frac{\partial }{\partial x^{j}}\left( \Delta _{\beta }F^{m}(\beta )\right) .
\notag
\end{gather}
Now, note that a Bochner type formula allows to compute the Laplacian of $%
e(\beta ;x)$ according to 
\begin{gather}
\Delta _{\beta }\left( e(\beta ;x)\right) =\Delta _{\beta }\left( \left|
\nabla F(\beta )\right| ^{2}\right)  \label{bochner} \\
=2\left| \nabla ^{2}F(\beta )\right| ^{2}+2Ric(\nabla F(\beta ),\nabla
F(\beta ))+2g\left( \nabla F(\beta ),\nabla \Delta _{\beta }F(\beta )\right) ,
\notag
\end{gather}
where the notation is as follows: 
\begin{gather}
\left| \nabla ^{2}F(\beta )\right| ^{2}\doteq g^{ik}(\beta )g^{jl}(\beta
)\delta _{rs}\nabla _{ij}^{2}F^{r}(\beta )\nabla _{kl}^{2}F^{s}(\beta ), \\
\nabla _{ij}^{2}F^{r}(\beta )\doteq \left[ \frac{\partial ^{2}}{\partial
x^{i}\partial x^{j}}F^{r}(\beta )-\Gamma _{\;\,ij}^{k}(g(\beta 
))\frac{\partial 
}{\partial x^{k}}F^{r}(\beta )\right] , \\
g\left( \nabla F(\beta ),\nabla \Delta _{\beta }F(\beta )\right) \doteq
g^{ik}(\beta )\delta _{rs}\nabla (\Delta _{\beta }F^{r}(\beta ))\nabla
_{k}F^{s}(\beta ), \\
Ric(\nabla F(\beta ),\nabla F(\beta ))\doteq \delta _{lm}\frac{\partial
F^{l}(\beta )}{\partial x^{i}}\frac{\partial F^{m}(\beta )}{\partial x^{j}}%
R^{ij}(\beta ).
\end{gather}
Equation (\ref{bochner}) implies that we can write (\ref{ener2}) as 
\begin{gather}
\frac{\partial }{\partial \beta }e(\beta ;x)=\delta _{lm}\frac{\partial
F^{l}(\beta )}{\partial x^{i}}\frac{\partial F^{m}(\beta )}{\partial x^{j}}%
\frac{\partial }{\partial \beta }g^{ij}(\beta )+\Delta _{\beta }\left(
e(\beta ;x)\right)  \label{evoleqn} \\
-2\left| \nabla ^{2}F(\beta )\right| ^{2}-2\delta _{lm}\frac{\partial
F^{l}(\beta )}{\partial x^{i}}\frac{\partial F^{m}(\beta )}{\partial x^{j}}%
R^{ij}(\beta ).  \notag
\end{gather}
This latter expression shows that, as expected, the 
$\beta $--evolution of $%
e(\beta ;x)$ is strongly influenced by the chosen evolution of \ $%
g^{ij}(\beta )$. However, such an evolution becomes particularly simple if
we choose the flow of metrics $g^{ij}(\beta )$ in a way that is strongly
reminiscent of the radial scaling properties of the metric (\ref{radscale}), 
\emph{viz.}, 
\begin{equation}
\frac{\partial }{\partial \beta }g^{ij}(\beta )=2R^{ij}(\beta ),
\end{equation}
or in terms of the covariant components $g_{ij}(\beta )$%
\begin{equation}
\frac{\partial }{\partial \beta }g_{ij}(\beta )=-2R_{ij}(\beta ).
\label{prosciut}
\end{equation}
Under such a choice (\ref{evoleqn}) reduces to 
\begin{equation}
\frac{\partial }{\partial \beta }e(\beta ;x)=\Delta _{\beta }\left( e(\beta
;x)\right) -2\left| \nabla ^{2}F(\beta )\right| ^{2},
\end{equation}
which implies that, as $\beta $ increases, the maximum of $e(\beta ;x)$ is
(weakly) decreasing. According to the geometrical meaning of \ $e(\beta ;x)$%
, (see (\ref{emeaning})), curvature inhomogeneities tend to decrease along
the flow (\ref{prosciut}), at least in a sufficiently small neighborhood (in
the $C^{k}$-topology) of \ $(B(p;r_{0}),g(\beta ))$.

\bigskip 

Obviously, these remarks are strongly reminiscent of the properties of the
Ricci flow on a Riemannian manifold $(\Sigma ,g)$, \emph{i.e.}, 
\begin{equation}
\left\{ 
\begin{tabular}{l}
$\frac{\partial }{\partial \eta }g_{ab}(\eta )=-2R_{ab}(\eta )$ \\ 
$g_{ab}(\eta =0)=g_{ab},$%
\end{tabular}
\right.   \label{Rflow}
\end{equation}
studied by Richard Hamilton and his co--workers in connection with an analytic
attempt to proving Thurston's geometrization conjecture. 
As is well--known,
the flow (\ref{Rflow}) is weakly--parabolic, and it is 
always solvable for
sufficiently small $\eta $. Moreover, it is such that any symmetries of $%
g_{ab}(\eta =0)=g_{ab}$ are preserved along the flow. The\ flow (\ref{Rflow}%
) may be reparametrized $\eta \rightarrow $ $\lambda $ by an $\eta $%
--dependent rescaling and by an $\eta $--dependent 
homothety $g_{ab}(\eta
)\rightarrow \widetilde{g}_{ab}(\lambda )$\ so as to preserve \ the original
volume of $(\Sigma ,g)$. In this latter case in place of (\ref{Rflow}) we
get the associated flow 
\begin{equation}
\left\{ 
\begin{tabular}{l}
$\frac{\partial }{\partial \lambda }\widetilde{g}_{ab}(\lambda )=-2%
\widetilde{R}_{ab}(\lambda )+\frac{2}{3}\left\langle \widetilde{R}(\lambda
)\right\rangle _{\Sigma (\lambda )}\widetilde{g}_{ab}(\lambda )$ \\ 
$\widetilde{g}_{ab}(\lambda =0)=g_{ab},$%
\end{tabular}
\right.   \label{unflow}
\end{equation}
whose global solutions (if attained) 
\begin{equation}
\overline{g}_{ab}\doteq \lim_{\lambda \rightarrow +\infty }\widetilde{g}%
_{ab}(\lambda )
\end{equation}
are constant curvature metrics.

\bigskip 

Thus, a Lagrangian picture shows in a very direct way that the natural flow $%
g_{ab}\rightarrow g_{ab}(\beta )$ providing the simultaneous  rescaling of
the matter distribution and the geometry is a (suitably normalized) Ricci flow.
Clearly, its normalization cannot be selected simply on the basis of
geometric convenience (\emph{e.g.}, fixed volume, as in the standard
Ricci--Hamilton flow(\ref{unflow})), and we can realize our matter averaging
program only if there exists a global solution $g_{ab}\rightarrow
g_{ab}(\beta )$, $0\leq \beta \leq \infty $ of the Ricci flow which
preserves the total matter content $M(\Sigma )$ of $(\Sigma ,g)$. Such a
problem together with the set up of a renormalized effective dynamics of 
smoothed out cosmological spacetimes is the headline of our forthcoming work.
\bigskip

\subsection*{\it Acknowledgements}

TB acknowledges support by 
the Tomalla Foundation, Switzerland, and the University of
Pavia during a visit in 2000.
Both of us acknowledge support by the 
Sonderforschungsbereich 375 `Astroparticle physics' during visits
to Ludwig--Maximilians--Universit\"at, Munich in 1999 and 2000.

\end{document}